\documentclass[cameraready]{Interspeech}
\usepackage{amsmath,amssymb,amsthm}
\usepackage{amsfonts} % mathbb
\usepackage{gensymb}  % for \degree
\usepackage{multicol}
\usepackage{multirow}
\usepackage{pifont}% http://ctan.org/pkg/pifont

\usepackage{xpatch,xcolor}
\usepackage{hyperref}

\usepackage{bbm}
\usepackage{booktabs}
\usepackage{makecell}
\title{Adaptive Federated Fine-Tuning of Self-Supervised Speech Representations}

\author[affiliation={1}, equalcontribution]{Xin}{Guo}
\author[affiliation={1}, equalcontribution]{Chunrui}{Zhao}
\author[affiliation={2}]{Hong}{Jia}
\author[affiliation={3}]{Ting}{Dang}
\author[affiliation={1}]{Gongping}{Huang}
\renewcommand{\authorsep}{\protect\\}
\author[affiliation={4}]{Xianrui}{Zheng}
\author[affiliation={4,5}]{Yan}{Gao}

\address{
    $^1$ Electronic Information School, Wuhan University, Wuhan, China \\
    $^2$ University of Auckland, Auckland, New Zealand \\
    $^3$ University of Melbourne, Melbourne, Australia \\
    $^4$ University of Cambridge, United Kingdom \\
    $^5$ Flower Labs, United Kingdom
}

\email{
xinguo19@whu.edu.cn,
chunruizhao@whu.edu.cn,
hong.jia@auckland.ac.nz,
ting.dang@unimelb.edu.au,
gongpinghuang@whu.edu.cn\\
xz396@cam.ac.uk,
yg381@cam.ac.uk
}


\usepackage{comment}

\begin{document}

\maketitle

\begin{abstract}
    Integrating Federated Learning (FL) with self-supervised learning (SSL) enables privacy-preserving fine-tuning for speech tasks. However, federated environments exhibit significant heterogeneity: clients differ in computational capacity, causing straggler effects under unified fine-tuning, while diverse downstream tasks require different representation depths, making full-model updates inefficient. To address these challenges, we propose an adaptive federated fine-tuning framework with early exits. Lightweight prediction heads are inserted at intermediate layers of the SSL backbone, allowing clients to terminate computation based on local constraints and task requirements. We further introduce a layer-wise, depth-aware partial aggregation strategy to better utilize representations from different network depths. Experiments show that the framework reduces edge overhead, supports heterogeneous hardware, and maintains competitive performance in resource-constrained federated environments.
\end{abstract}

\section{Introduction}

Self-supervised learning (SSL) approaches for speech, such as Wav2Vec 2.0 and HuBERT, substantially improve feature extraction by leveraging large-scale unlabeled data during pre-training.~\cite{baevski2020wav2vec,hsu2021hubert,schneider2019wav2vec,devlin2019bert}. Through contrastive learning or masked prediction, these models learn universal speech representations capturing both acoustic and semantic information. With only limited labeled data for fine-tuning, they achieve strong performance on downstream tasks including automatic speech recognition, speaker identification, and emotion recognition~\cite{graves2013speech,snyder2018x,schuller2009interspeech}. 
However, conventional centralized fine-tuning of speech SSL models requires aggregating data at a central server. This approach is often impractical in scenarios involving sensitive data, such as personal assistants and mobile devices. Consequently, it raises compliance concerns under increasingly stringent privacy regulations, such as the General Data Protection Regulation~\cite{voigt2017eu}. 
Federated learning (FL) addresses this issue by enabling collaborative model training while keeping raw speech data local~\cite{mcmahan2017communication}. By aggregating model updates rather than raw data, FL reduces privacy risks while leveraging distributed data resources~\cite{yang2019federated,gao2022federated,gao2022end}.

Despite the promise of combining SSL fine-tuning and federated learning, practical deployment faces significant heterogeneity challenges~\cite{li2020federated}. First, system heterogeneity affects training efficiency. Federated networks consist of devices with vastly different computational capabilities, from high-performance servers to low-power IoT devices~\cite{lim2020federated}. Large SSL models with deep Transformer architectures~\cite{vaswani2017attention} impose substantial computational and memory burdens on resource-constrained devices during full fine-tuning, often resulting in high latency, energy consumption, or even memory overflow. 
Second, heterogeneity of downstream tasks makes a unified fine-tuning strategy suboptimal. Speech representations are inherently hierarchical: shallow layers capture acoustic patterns sufficient for simple classification tasks such as keyword spotting, while deeper layers encode high-level semantic and contextual information required for complex tasks like automatic speech recognition~\cite{pasad2021layer}. Existing federated fine-tuning approaches typically activate the entire backbone for all tasks~\cite{horvath2021fjord}, leading to unnecessary computational overhead~\cite{teerapittayanon2016branchynet}

To address these challenges, we propose an adaptive early-exit fine-tuning framework for heterogeneous federated environments, designed to accommodate diverse downstream tasks~\cite{teerapittayanon2016branchynet}.
We transform a static deep SSL model into an elastic multi-branch architecture by inserting task-specific prediction heads at intermediate layers~\cite{kaya2019shallow}, enabling early termination during both training and inference. We further design a dual-adaptive strategy that considers both hardware constraints and task requirements. Each client determines its maximum trainable depth based on local computational resources and selects an appropriate exit layer according to task complexity. This design allows low-resource devices to contribute shallow-layer updates while enabling high-resource devices to optimize deeper representations when necessary. To aggregate parameters from models with varying depths, we introduce a layer-wise partial aggregation mechanism that selectively combines depth-weighted updates from clients participating in each layer~\cite{diao2021heterofl}. Extensive experiments across five downstream tasks demonstrate that the proposed framework enables robust and efficient training while maintaining strong performance under heterogeneous conditions.
\vspace{-5pt}
\section{Method}
\begin{figure*}[t] 
  \centering
  \includegraphics[width=0.9\linewidth]{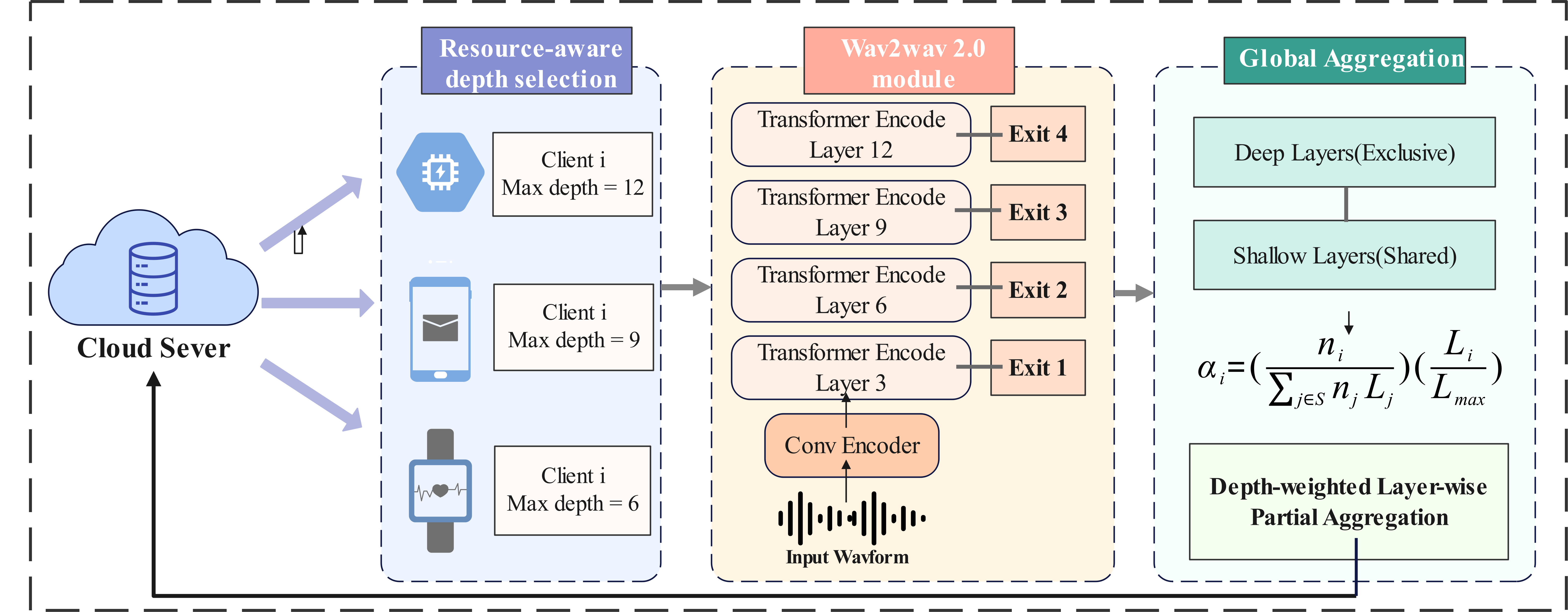} 
  \caption{Overview of the proposed Adaptive Federated Learning framework. Heterogeneous clients dynamically select the training depth based on local resources. The server employs a Depth-weighted Layer-wise Partial Aggregation strategy, where deeper layers are updated exclusively by high-resource clients.}
  \label{fig:fl}
\end{figure*}

To address the challenges of system and task heterogeneity in federated SSL model fine-tuning, we propose an adaptive framework as illustrated in Figure~\ref{fig:fl}.
The framework consists of three components: (1) resource-aware depth selection by clients, (2) an elastic multi-exit backbone model, and (3) a depth-weighted layer-wise partial aggregation mechanism on the server.

\subsection{Multi-Exit Elastic Backbone}

The proposed framework is built upon a pre-trained Wav2Vec~2.0 model~\cite{baevski2020wav2vec} and aims to enable adaptive fine-tuning in federated environments through a structured early-exit design. Wav2Vec~2.0 is a self-supervised speech representation model consisting of a multi-layer convolutional feature encoder followed by a 12-layer Transformer-based contextual encoder, which extracts rich acoustic representations directly from raw waveforms. To accommodate multi-task requirements and resource heterogeneity in federated settings, we depart from the standard single-output architecture and transform the backbone into a multi-exit elastic architecture~\cite{xin2020deebert}. Specifically, we attach independent prediction heads to a selected set of intermediate Transformer layers, denoted as $\mathcal{E}$. This modification converts the original monolithic deep network into a hierarchical feature extractor with progressively increasing representational depth, thereby providing the structural foundation for adaptive and elastic training across heterogeneous clients.

% To accommodate diverse downstream speech tasks, the prediction heads attached to each exit point are tailored according to the task category:

% \begin{itemize}
%     \item \textbf{Classification tasks}, including speaker identification (SID), emotion recognition (ER), and keyword spotting (KWS): 
%     The prediction head incorporates a temporal pooling module to aggregate variable-length frame-level representations into a fixed-dimensional utterance-level embedding, such as an x-vector representation~\cite{snyder2018x}. 
%     The aggregated embedding is subsequently processed by fully connected layers to perform category prediction.

%     \item \textbf{Sequence generation tasks}, such as automatic speech recognition (ASR): 
%     The prediction head is composed of a linear projection layer followed by a Connectionist Temporal Classification (CTC) objective layer~\cite{graves2006connectionist}. 
%     This structure enables direct decoding from frame-level representations while preserving temporal alignment and projecting the features into the target vocabulary space.
% \end{itemize}

\subsection{Resource and Task-Aware Local Training}

To address client heterogeneity in federated environments, we design an adaptive local training scheme. 
Prior to each communication round, each client dynamically determines its maximum trainable network depth, denoted as $L_{\max} \in \mathcal{E}$, based on its local hardware resources (e.g., memory capacity and computational capability) and the complexity of the assigned task.

This mechanism considers two complementary factors. 
First, from a resource-awareness perspective, a safe upper bound on network depth is imposed for resource-constrained devices to prevent memory overflow and excessive computational burden. 
Second, from a task-awareness perspective, relatively shallow exit points are strategically selected for less complex tasks, such as keyword spotting, in order to avoid redundant computation.

During local training, forward propagation is executed only up to layer $L_{\max}$, where the task-specific loss is computed. 
Backpropagation updates parameters exclusively within layers up to $L_{\max}$, while deeper layers remain frozen. 
This dynamic pruning strategy enables all clients to contribute their local knowledge in a computationally efficient manner~\cite{han2021dynamic}, without being constrained by the weakest devices in the network.

\subsection{Layer-wise Depth-aware Partial Aggregation}

Heterogeneous training depths across clients cause dimensional mismatch in updates, making standard FedAvg inapplicable. 
We adopt a layer-wise partial aggregation with depth-aware weighting.

Instead of averaging the entire backbone, the server aggregates each Transformer layer independently. 
For layer $l$, only clients that trained it contribute. 
This allows lower layers to be collaboratively optimized even when higher layers are trained by only a subset of clients. 
Aggregation weights are proportional to local data size $n_i$ and maximum trained depth $L_i$:

\begin{equation}
\mathbf{w}_{l}^{(t+1)} =
\frac{
\sum_{i \in \mathcal{S}_l} (n_i \cdot L_i) \mathbf{w}_{l,i}^{(t+1)}
}{
\sum_{j \in \mathcal{S}_l} (n_j \cdot L_j)
},
\end{equation}

where $\mathcal{S}_l$ is the set of clients training layer $l$, and $\mathbf{w}_{l,i}^{(t+1)}$ are their local updates. 
This induces a pyramid-shaped optimization: lower acoustic layers are updated by all clients for statistical robustness, while higher semantic layers are refined by clients training deeper segments. 
Decoupling depth from device capability mitigates the straggler effect and stabilizes convergence in heterogeneous federated speech learning.

\section{Experimental Setup}

\subsection{Datasets and Downstream Tasks}

To comprehensively evaluate the effectiveness of the proposed federated early-exit fine-tuning framework based on Wav2Vec~2.0 Base, we follow the task categorization protocol of the SUPERB benchmark and select five representative downstream speech tasks~\cite{yang2021superb}. 
These tasks span multiple aspects of speech understanding, including speech content recognition, speaker modeling, and paralinguistic analysis.

% \begin{itemize}
%     \item \textbf{Speech content recognition}: 
%     This category includes ASR and KWS. 
%     For ASR, experiments are conducted on the LibriSpeech dataset~\cite{panayotov2015librispeech} using the CTC objective, and performance is evaluated in terms of word error rate (WER). 
%     For KWS, we use the Google Speech Commands dataset~\cite{tang2018deep}, considering both the full 35-class set and a subset of 12 target commands, and report classification accuracy.

%     \item \textbf{Paralinguistic and speaker-related tasks}: 
%     This category includes ER, SID, and ASV. 
%     ER is evaluated on the IEMOCAP dataset~\cite{busso2008iemocap} under a four-class emotion classification setting. 
%     SID and ASV are conducted on the VoxCeleb1 dataset~\cite{nagrani2017voxceleb}, where multi-class classification accuracy and equal error rate (EER) are reported, respectively.
% \end{itemize}

\vspace{1mm}
\noindent\textbf{Speech content recognition}: 
This category includes \textbf{Automatic Speech Recognition (ASR)} and \textbf{Keyword Spotting (KWS)}, while adopting different prediction heads and training objectives.
For ASR, the frame-level representations are further fed into a sequence modeling module, then projected to the vocabulary space by a linear layer, and trained with the \textbf{Connectionist Temporal Classification (CTC)} objective~\cite{graves2006connectionist}. 
Experiments are conducted on the LibriSpeech dataset~\cite{panayotov2015librispeech}, and performance is evaluated in terms of word error rate (WER).
For KWS, the model aggregates frame-level representations into a fixed-dimensional embedding via statistics pooling~\cite{snyder2018x}, and performs category prediction using a linear layer followed by a softmax. 
We evaluate KWS on the Google Speech Commands dataset~\cite{speechcommandsv2}, considering both the full 35-class set and a 12-command subset, and report the classification error rate.

\vspace{1mm}
\noindent\textbf{Paralinguistic and speaker-related tasks}: 
This category includes \textbf{Emotion Recognition (ER)}, \textbf{Speaker Identification (SID)}, and \textbf{Automatic Speaker Verification (ASV)}.
For ER, the model similarly adopts an utterance-level classification head with ``statistics pooling + classifier'', aggregating frame-level representations into a fixed-dimensional utterance embedding for emotion classification. 
ER is evaluated on the IEMOCAP dataset~\cite{busso2008iemocap} under a four-class setting.
For SID and ASV, experiments are conducted on the VoxCeleb1 dataset~\cite{nagrani2017voxceleb}. 
Specifically, SID takes utterance-level speaker embeddings as input and predicts speaker identities via a classifier, reporting the multi-class classification error rate.
ASV reuses the same utterance-level embeddings for speaker verification and reports the equal error rate (EER).

The diversity of these tasks ensures comprehensive evaluation across varying levels of semantic complexity and representation depth requirements.

\subsection{Model Architecture and Early-Exit Configuration}

We adopt Wav2Vec~2.0 Base as the shared backbone model in the federated learning framework. 
The model consists of a convolutional feature encoder followed by a 12-layer Transformer-based contextual encoder, which extracts hierarchical semantic representations from raw speech waveforms. 
All experiments are initialized using model parameters pre-trained on the 100-hour subset of the LibriSpeech corpus.

For downstream tasks, the frame-level representations produced by the encoder are first aggregated via statistical pooling to obtain utterance-level embeddings~\cite{cai2018exploring}, which are then passed through a linear projection layer to generate task-specific predictions.
To evaluate the effectiveness of elastic depth adaptation, early-exit points are introduced at the 3rd, 6th, 9th, and 12th Transformer layers, forming the exit set
$\mathcal{E} = \{3, 6, 9, 12\}$.
All configurations share identical pre-trained initialization and differ only in the effective encoder depth, resulting in varying computational complexity and representational capacity. 
Shallower configurations substantially reduce computational overhead and are therefore suitable for resource-constrained clients, whereas deeper configurations provide stronger contextual modeling capability and benefit more complex tasks.

This depth-based structural design establishes a unified framework for hierarchical computation and heterogeneous collaborative training in federated environments, enabling a controllable and scalable trade-off between model performance and computational cost.

\subsection{Federated Learning Environment and Heterogeneity}

All experiments are implemented using the Flower federated learning framework in conjunction with the SpeechBrain speech processing toolkit~\cite{beutel2020flower,speechbrain_v1,speechbrain}. 
We consider three scenarios: centralized training, homogeneous federated learning, and heterogeneous federated learning.

\textbf{Centralized Training (Upper Bound).}
Data from all clients are aggregated for end-to-end fine-tuning. We conduct experiments for encoder depths at the 3rd, 6th, 9th, and 12th Transformer layers. This provides an ideal performance reference without privacy or communication constraints.

\textbf{Homogeneous Federated Learning.}
All clients adopt identical model depths (e.g., all train up to the 6th). This setting investigates convergence behaviors and quantifies the performance gap between distributed and centralized training.

\textbf{Heterogeneous Federated Learning.}
To emulate realistic edge environments, we introduce heterogeneity along two dimensions:
\textit{1) System Heterogeneity:}  
Clients are divided into two groups, each accounting for 50\%. 
Low-resource clients simulate IoT or mobile devices and are restricted to shallow layers (e.g., the 3rd Transformer layer). High-resource clients simulate high-performance servers and train up to the task-specific optimal layer, rather than a fixed 12-layer model.
\textit{2) Data Heterogeneity:}
For the KWS, ER, and ASR tasks, data are partitioned strictly according to speaker identity, such that each client contains speech samples from a limited subset of speakers, forming a highly challenging non-IID setting~\cite{zhang2023fedaudio,zhu2022decouplefl_asr}. 
For SID and ASV tasks, data are randomly shuffled across clients (IID partitioning) to better align with task characteristics.

In all federated experiments, a fraction of clients is randomly sampled in each communication round. 
Each selected client performs local optimization for 1--3 epochs before uploading updated parameters to the server. 
The number of communication rounds is fixed for each task to ensure stable convergence. KWS is trained for 1000 rounds due to its slower convergence under non-IID settings. ASR is trained for 100 rounds. SID is trained for 200 rounds to obtain sufficiently discriminative speaker representations. ER is trained for 50 rounds as it converges relatively faster.
ASV is not trained in a federated manner. Instead, it is evaluated using the global model obtained from the SID task after 200 communication rounds.
After each aggregation step, the global model is evaluated on a centralized test set to monitor performance progression.

\begin{table*}[t]
\centering
\caption{Performance comparison under different model depths. 
All results are reported in percentage (\%). 
KWS, ER, and SID are evaluated using classification error rate; 
ASR is evaluated using WER; 
and ASV is evaluated EER. }
\label{tab:layer_performance}

\begin{tabular}{l|l|cccc|cccc}
\toprule
\multirow{2}{*}{Tasks} 
& \multirow{2}{*}{Metric}
& \multicolumn{4}{c}{Centralised} 
& \multicolumn{4}{c}{Federated Learning} \\
\cmidrule(lr){3-6} 
\cmidrule(lr){7-10}
& & 12 layer & 9 layer & 6 layer & 3 layer 
& 12 layer & 9 layer & 6 layer & 3 layer \\
\midrule

KWS (12 commands) 
& Error Rate(\%) $\downarrow$
& 5.23 & 5.31 & 4.93 & 5.86 
& 20.30 & 18.80 & \textbf{17.50} & 24.90 \\

KWS (35 commands) 
& Error Rate(\%) $\downarrow$
& 16.00 & 16.80 & 16.80 & 16.80 
& 15.30 & 12.40 & \textbf{10.70} & 21.70 \\

ASR (test-clean) 
& WER(\%) $\downarrow$
& 9.00 & 6.46 & 10.00 & 14.83 
& 12.88 & \textbf{8.81} & 13.26 & 21.73 \\

ASR (test-other) 
& WER(\%) $\downarrow$
& 23.27 & 15.92 & 24.59 & 37.63 
& 28.44 & \textbf{18.41} & 28.70 & 45.61 \\

ER 
& Error Rate(\%) $\downarrow$
& 41.00 & 40.10 & 38.00 & 38.80 
& 39.50 & \textbf{35.70} & 36.60 & 37.70 \\

SID 
& Error Rate(\%) $\downarrow$
& 12.90 & 11.10 & 10.30 & 7.00 
& 19.00 & \textbf{17.30} & 20.10 & 17.70 \\

ASV 
& EER(\%) $\downarrow$
& 6.80 & 5.40 & 5.40 & 4.50 
& \textbf{13.40} & 20.40 & 15.64 & 20.50 \\

\bottomrule
\end{tabular}
\end{table*}

\begin{table}[t]
\centering
\caption{Performance comparison across different federated settings. 
Three experiments are compared: Homogeneous FL (Homo-FL), all clients with the optimal exit depth for the task. 
Heterogeneous FL with standard FedAvg aggregation, and Heterogeneous FL with layer-wise partial aggregation. All results are reported in percentage (\%). }
\label{tab:heterogeneous}

\begin{tabular}{l|ccc}
\toprule
Tasks & Homo-FL & FedAvg & Layer-wise \\
\midrule

KWS (12 commands) & 17.50 & 18.40 & 17.60 \\
KWS (35 commands) & 10.70 & 14.60 & 13.40 \\
ASR (test-clean) & 8.81 & 9.21 & \textbf{8.79} \\
ASR (test-other) & 18.41 & 19.20 & \textbf{18.40} \\
ER & 35.70 & 35.00 & \textbf{34.50} \\
SID & 17.30 & 17.50 & \textbf{15.30} \\
ASV & 20.40 & 12.15 & \textbf{10.93} \\

\bottomrule
\end{tabular}
\vspace{-10pt}
\end{table}
\section{Experimental Results}

\subsection{Layer-wise Performance Analysis under Centralized and Federated Settings}

We evaluate layer-wise performance under centralized training and homogeneous federated learning (FL). Results show that optimal exit layers vary across tasks, indicating the full 12-layer representation is not always optimal.

Table~\ref{tab:layer_performance} summarizes performance across multiple speech tasks. Overall, ASR exhibits a consistent preference for the 9th layer under both training paradigms, with the 9th-layer exit achieving the best performance in both centralized and federated learning (FL). A plausible reason is that ASR relies more on higher-level context modeling and semantic integration, for which mid-to-deep representations are better suited for sequence-to-text recognition~\cite{pasad2021layer}. In contrast, classification tasks show more diversified layer preferences in FL: the optimal exits for KWS-12 and KWS-35 both occur at the 6th layer, whereas ER performs best at the 9th layer. Intuitively, KWS depends more on stable local acoustic cues and short-term patterns, where intermediate representations are often sufficiently discriminative and pushing to deeper, more semantic representations does not necessarily help; ER, however, benefits from richer prosodic and higher-level expressive information and therefore prefers deeper representations. On IEMOCAP, the 9th-layer FL model outperforms the best centralized result, which may be attributed to the implicit regularization effect introduced by federated averaging. For speaker identification (SID), the optimal FL exit is also at the 9th layer, suggesting that under non-IID federated training, mid-to-deep layers provide more robust aggregation of speaker-related characteristics. By contrast, the optimal exit layers in centralized training are more scattered, for example, ER and KWS-12 tend to favor the 6th layer, while SID achieves the best result at the shallower 3rd layer.

\vspace{-5pt}
\subsection{Layer-wise Partial Aggregation for Heterogeneous Clients}

We conduct two federated learning experiments for comparison. First, clients are split into shallow and deep groups, but updates are still aggregated using conventional FedAvg. In the second, the same heterogeneous depth configuration is combined with our proposed layer-wise partial aggregation mechanism (Layer-wise Aggregation), where each Transformer layer is aggregated independently with depth-aware weighting.

As shown in Table~\ref{tab:heterogeneous}, Homogeneous FL (where all clients use the task-specific optimal exit layer) serves as a reference baseline. Under the heterogeneous-depth setting, introducing depth-aware weighting into the layer-wise aggregation further improves overall performance, and even surpasses the homogeneous FL baseline on some tasks (e.g., ASR and ER). This suggests that depth-aware layer-wise aggregation not only mitigates the instability and performance degradation caused by heterogeneity, but can also yield additional gains by integrating updates from clients of different depths in a more principled manner. More importantly, this strategy provides a practical solution for hardware-heterogeneous federated learning, enabling resource-constrained shallow clients and more capable deep clients to collaborate within a unified training framework, thereby offering a more flexible trade-off between performance and computation/communication cost.

% Additionally, for the ER task, we further examine the impact of different adaptive federated aggregation optimizers on training outcomes. Specifically, we compare three aggregation strategies, FedAdagrad, FedAdam, and FedYogi, all of which employ per-parameter adaptive updates to better accommodate heterogeneous client updates. Table~\ref{tab:er_optimizers} summarizes the ER performance under these strategies. Overall, the choice of aggregation strategy leads to noticeable differences; under our experimental setting, FedAdam provides a relatively better trade-off between classification accuracy and training stability.

% \begin{table}[t]
% \centering
% \caption{ER performance under different adaptive federated optimizers. 
% Results are reported as Equal Error Rate (EER, \%).}
% \label{tab:er_optimizers}
% \begin{tabular}{lccc}
% \toprule
% Optimizer & FedAdagrad & FedAdam & FedYogi \\
% \midrule
% ER  &  &  &  \\
% \bottomrule
% \end{tabular}
% \vspace{-10pt}
% \end{table}

% \vspace{-5pt}
\subsection{Memory Cost under Varying Model Depth}

% \begin{table}[!t]
% \centering
% \caption{Memory cost (MB) as a function of Transformer depth across four tasks.
% Measurements are conducted with fixed batch size using dummy inputs.}
% \label{tab:memory_cost}
% \begin{tabular}{lcccc}
% \toprule
% Tasks & 3 Layers & 6 Layers & 9 Layers & 12 Layers \\
% \midrule
% KWS & 319.02 & 400.14 & 481.25 & 562.36 \\
% ER  & 1328.76 & 1409.88 & 1490.99 & 1572.11 \\
% ASR & 2293.20 & 2374.31 & 2455.43 & 2536.54 \\
% SID & 750.14 & 831.47 & 912.37 & 993.70 \\
% \bottomrule
% \end{tabular}
% \vspace{-15pt}
% \end{table}
\begin{table}[!t]
\centering
\caption{Memory cost (MB) as a function of Transformer depth across four tasks. Measurements are conducted using speech inputs with a fixed input length and batch size.}
\label{tab:memory_cost}
\begin{tabular}{l|cccc}
\toprule
\multirow[c]{2}{*}{\strut Tasks} & \multicolumn{4}{c}{Memory cost (MB)} \\
\cmidrule(lr){2-5}
& 3 Layers & 6 Layers & 9 Layers & 12 Layers \\
\midrule
KWS & 319.02  & 400.14  & 481.25  & 562.36 \\
ER  & 1328.76 & 1409.88 & 1490.99 & 1572.11 \\
ASR & 2293.20 & 2374.31 & 2455.43 & 2536.54 \\
SID & 750.14  & 831.47  & 912.37  & 993.70 \\
\bottomrule
\end{tabular}
\vspace{-10pt}
\end{table}
To better understand the system behavior under resource constraints,
we analyze the GPU memory consumption with different Transformer depths.
As shown in Table~\ref{tab:memory_cost}, memory usage increases approximately linearly as additional Transformer layers are included.
Across all four tasks, reducing the depth from 12 to 3 layers yields memory reductions of 43.27\% (KWS), 24.51\% (SID), 15.48\% (ER), and 9.59\% (ASR), respectively.
These results demonstrate that depth scaling offers a predictable and effective mechanism for controlling client-side memory consumption, which is particularly important in federated settings with heterogeneous hardware constraints.

\section{Conclusion}

This paper addresses the problem of efficient fine-tuning of self-supervised speech models in heterogeneous federated environments. We propose an adaptive federated training framework based on an early-exit mechanism. By constructing a multi-exit elastic backbone and incorporating resource- and task-aware local training strategies together with a depth-weighted layer-wise partial aggregation scheme, the proposed method enables collaborative optimization among clients with diverse computational capabilities. Experimental results show that intermediate-layer representations consistently outperform the full-depth model across multiple speech tasks, demonstrating that early-exit structures can improve efficiency while maintaining or even enhancing performance. In heterogeneous federated settings, the framework achieves stable convergence and effectively mitigates the straggler effect. Overall, this work provides a scalable and efficient solution for deploying large-scale self-supervised speech models in resource-constrained environments.\looseness=-1

\section{Generative AI Use Disclosure}
During the preparation of this work, the authors used ChatGPT and Gemini in order to polish the language and improve readability. After using these tools, the authors reviewed and edited the content as needed and take full responsibility for the content of the publication.

\bibliographystyle{IEEEtran}
\bibliography{mybib}

\end{document}